\documentclass[apj]{emulateapj}
\begin{document}

\title{Chemical Composition of Faint ($I\sim21$ mag) Microlensed Bulge Dwarf OGLE-2007-BLG-514S}
\shorttitle{Chemical Compostion of OGLE-2007-BLG-514S}
\author{
Courtney~R.~Epstein\altaffilmark{1},
Jennifer~A.~Johnson\altaffilmark{1},
Subo Dong\altaffilmark{1,}\altaffilmark{2,}\altaffilmark{3},
Andrzej Udalski\altaffilmark{4},
Andrew Gould\altaffilmark{1,}\altaffilmark{2}, and
George Becker\altaffilmark{5,}\altaffilmark{6}
}
\altaffiltext{1}
{Department of Astronomy, Ohio State University,
140 W.\ 18th Ave., Columbus, OH 43210, USA;
epstein,jaj,gould@astronomy.ohio-state.edu}
\altaffiltext{2}
{Microlensing Follow-Up Network ($\mu$FUN)}
\altaffiltext{3}
{Institute for Advanced Study, School of Natural Sciences, Einstein Drive, Princeton, NJ, 08540; dong@ias.edu}
\altaffiltext{4}
{Warsaw University Observatory,
Warszawa, Poland; udalski@astrouw.edu.pl}
\altaffiltext{5}
{Carnegie Observatories,
813 Santa Barbara Street, Pasadena, CA 91101, USA
}
\altaffiltext{6}
{Fellow, Kavli Institute for Cosmology and Institute of Astronomy, Madingley Road, Cambridge CB3 0HA, UK; gdb@ast.cam.ac.uk
}

\begin{abstract}
We present a high-resolution spectrum of a microlensed G dwarf in the Galactic bulge with spectroscopic temperature $T_\mathrm{eff} = 5600 \pm 180$ K. This $I \sim 21$ mag star was magnified by a factor ranging from 1160 to 1300 at the time of observation. Its high metallicity ($\mathrm{[Fe/H]} = 0.33 \pm 0.15$) places this star at the upper end of the bulge giant metallicity distribution. Using a K-S test, we find a 1.6\% probability that the published microlensed bulge dwarfs share an underlying distribution with bulge giants, properly accounting for a radial bulge metallicity gradient. We obtain abundance measurements for 15 elements and perform a rigorous error analysis that includes covariances between parameters. This star, like bulge giants with the same metallicity, shows no alpha enhancement. It confirms the chemical abundance trends observed in previously analyzed bulge dwarfs. At supersolar metallicities, we observe a discrepancy between bulge giant and bulge dwarf Na abundances.
\end{abstract}

\section{Introduction}
Baade's discovery (1946, 1958) that the central regions of the Galaxy, like other Sb spirals, contained population II stars sparked intense interest in the Galactic bulge. Because it is the closest galactic spheroid to the Sun, we can study the Galactic bulge in unique detail and use it to better understand galaxy formation and evolution.

Proposed senarios for building bulges include mergers and secular dynamical evolution (e.g.\ \citealt{kormendy:04}). Discriminating between bulge formation theories hinges on age determinations of bulge stars, a process complicated by crowding, contamination by foreground stars, reddening, and dispersion in both metallicity and distance along the line of sight. Comparing main-sequence photometry of bulge globular clusters with halo globular clusters demonstrated that they are coeval \citep{ortolani:95}. Examining the color-magnitude diagrams of both bulge fields and clusters shows that the bulge clusters are representative of the general bulge population, the majority of which are old stars ($\sim10$ Gyr) (e.g.\ Ortolani et al.\ 1995; Zoccali et al.\ 2003). Although \citet{feltzing:00} agree that the bulge is mostly old, their interpretation of the photometry permits a young metal-rich population of stars.

Evidence for a short star formation phase comes from the chemical composition of bulge stars. The $\alpha$-elements (O, Mg, Si, Ca, and Ti) are created in the evolution and explosion of massive stars as core collapse supernovae (SNcc) and almost immediately recycled in the interstellar medium (ISM) (e.g.\ \citealt{woosley:95}). Because of their longer progenitor lifespan, Type Ia supernovae (SNIa) do not enrich the ISM with substantial amounts of iron until 1-1.5 Gyr after star formation begins \citep{matteucci:09}. The delayed SNIa iron contribution dilutes the SNcc-enhanced $\alpha$/Fe ratio until it reaches solar levels \citep{tinsley:79}. Since bulge giants maintain an elevated $\alpha$/Fe ratio to high metallicity, the bulge must have evolved faster than the solar neighborhood and undergone an intense burst of star formation early in the history of the galaxy (e.g.\ \citealt{matteucci:90,mcwilliam:94}).

Within this framework, the $\alpha$-elements exhibit different behavior. The explosive $\alpha$-elements Si, Ca, and Ti appear to track each other \citep{fulbright:07}. Surprisingly, the hydrostatic $\alpha$-elements O and Mg do not, even though both are made in SNcc progenitors and released in SN explosions. In particular, Mg remains enhanced in the bulge to supersolar metallicities \citep{mcwilliam:94,fulbright:07}. The bulge O abundance closely tracks the abundance trends in disk stars and begins declining around [Fe/H]$\approx-0.3$ dex (e.g.\ \citealt{melendez:08}). \citet{mcwilliam:08} reconciles these trends by invoking Wolf-Rayet winds in metal-rich, massive stars to produce lower mass SNcc progenitors, which \citet{woosley:95} predict yield less oxygen.

Neutron-capture elements also have the potential to act as clocks. AGB stars are associated with the production of $s$-process elements, while supernovae may be responsible for $r$-process elements. Little information about neutron-capture elements is available in the bulge since the absorption lines are concentrated in the crowded blue portion of the spectrum. Nevertheless, \citet{mcwilliam:04} observed an enhanced Eu abundance and tentatively suggest a behavior similar to Ca. \citet{mcwilliam:94} find a subsolar $s$-process abundance, reminiscent of old, metal-poor halo stars, indicating that the $r$-process dominates in the bulge.

Light elements, like Na, contain important information about the chemical evolution of the bulge. Na production is sensitive to the neutron excess \citep{arnett:71} and its yield is therefore dependent on the metallicity of its SNcc progenitor. Although typically interpreted as SNcc ejectum, Na can also be synthesized in envelope burning of AGB stars. \citet{lecureur:07} report a flat Na abundance trend in the bulge below solar metallicity, after which it increases abruptly with a dispersion at high-metallicities exceeding measurement error. Lecureur et al.\ determine that the elevated Na abundance cannot be explained by the mild internal mixing seen in their sample of giants and therefore must reflect the composition of the ISM at its formation.

Our knowledge of the bulge's chemical evolution comes from spectroscopic studies of K and M giants because they are typically the only stars accessible to high-resolution observations due to the far distance and the high degree of interstellar extinction toward the bulge. However, large surveys designed to identify and follow up microlensing events, such as the Optical Gravitional Lens Experiment\footnote{http://ogle.astrouw.edu.pl/ogle3/ews/ews.html} (OGLE) and the Microlensing Observations in Astrophysics (MOA) collaboration\footnote{www.massey.ac.nz/$\sim$iabond/alert/alert.html}, have made possible the high-resolution spectroscopic study of otherwise unobservable bulge dwarfs. \citet{lennon:96} performed the first spectroscopic abundance determination for a microlensed bulge dwarf, 96-BLG-3, detected by the MACHO collaboration. Based on low-resolution data, they estimated a metallicity of between 0.3 and 0.6 dex. \citet{minniti:98} obtained a high-resolution spectrum of a dwarf using Keck. \citet{cavallo:03} presented a preliminary abundance analysis of six microlensed bulge stars, including the Minniti et al.\ star and two other dwarfs.

There are seven published abundances analyses of microlensed bulge dwarfs based on high-resolution data: OGLE-2006-BLG-265S \citep{johnson:07}, OGLE-2007-BLG-349S \citep{cohen:08}, MOA-2006-BLG-099S \citep{johnson:08}, OGLE-2008-BLG-209S \citep{bensby209:09}, MOA-2008-BLG-310S and MOA-2008-BLG-311S \citep{cohen:09}, and OGLE-2009-BLG-076S \citep{bensby76:09}. Although OGLE-2008-BLG-209S is actually  a sub-giant, we will refer to these seven stars collectively as bulge dwarfs. Additionally, \citet{bensbyIAU:09} announced that high-resolution spectra have been obtained for six more bulge dwarfs, including a reanalysis of the Cavallo et al.\ dwarfs, whose detailed abundances will appear in Bensby et al. (in prep.) and Johnson et al. (in prep.).

Studying dwarfs offers several advantages. Dwarfs are subject to different systematic effects than the giants. Mixing on the red giant branch (RGB) can destroy some elements and dredge up others, altering a star's surface composition. Some elements that are difficult to see in giants, like S and Zn, can be measured in dwarfs since their hotter temperatures strengthen these atomic lines while weakening CN. Additionally, we can compare the metallicity distribution function (MDF) for dwarfs and giants.

The first-observed high-magnification bulge dwarf with a high-resolution spectrum had a surprisingly high metallicity of 0.56 dex \citep{johnson:07}. Subsequent observations revealed more metal-rich dwarfs indicating that high metallicities are common in bulge dwarfs. By contrast, the bulge giant MDF peaks just below solar metallicity (e.g.\ \citealt{fulbright:07,cunha:06,rich:05,rich:07,lecureur:07,zoccali:03,zoccali:08}). \citet{cohen:08} suggest that high-metallicity stars experience a sufficiently high mass-loss rate to cause them to peel off the RGB before undergoing a He flash and become He white dwarfs (WDs). \citet{zoccali:08} counter that this scenario would produce a decline in the RGB luminosity function, which is not observed in the \citet{zoccali:03} data. OGLE-2008-BLG-209S, the first bulge sub-giant with a high-resolution spectrum, was found to have a subsolar metallicity, consistent with the giants also located in Baade's window \citep{bensby209:09}. The detection of metal-poor dwarf OGLE-2009-BLG-076S opposes the idea of a strong selection effect against metal-poor microlensed stars.

Here we present both the high-magnification bulge dwarf MDF and a detailed chemical abundance analysis of the eighth bulge dwarf, OGLE-2007-BLG-514S.

\section{Observations and Data Reduction}\label{sec:obs}

\begin{figure}
\plotone{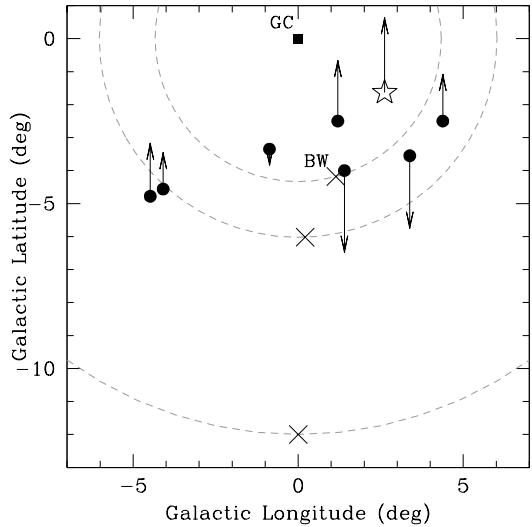}
\caption{Distribution in Galactic latitude and longitude of dwarf stars in the bulge that have published high-resolution spectral analyses (solid circles). A star marks the location of OGLE-2007-BLG-514S. The heliocentric radial velocity for each star is indicated by an arrow where up corresponds to a positive velocity and the vectors are scaled so that one degree in length is 70 km s$^{-1}$. The crosses indicate the fields from \citet{zoccali:08}, including Baade's window (BW), whose MDF is discussed in \S\ref{sec:MDF}. The dashed circles mark the locus where the projected distance from the Galactic Center (GC) equals that of a Zoccali et al. field.}\label{fig:location}
\end{figure}

OGLE-2007-BLG-514S is located toward the Galactic bulge (J2000 R.A. $= 17^\mathrm{h}58^\mathrm{m}3\fs09$, Dec. $= -27\degr31\arcmin05\farcs7$; $l = 2.62$, $b=-1.63$). Figure \ref{fig:location} gives the position of this star and the other seven bulge dwarfs. Of these stars, this dwarf lies the closest to the plane of the Galaxy.

OGLE-2007-BLG-514S's dereddened Johnson-Cousins color and magnitude relative to the red clump (see \S\ref{sec:photometric parameters}) are consistent with it being a main sequence turnoff star at the distance of the red clump. Alerted to this situation, Michael Rauch and George Becker took three high-resolution spectra of OGLE-2007-BLG-514S with the Magellan Inamori Kyocera Echelle (MIKE) double spectrograph mounted on the Clay telescope at Las Campanas \citep{bernstein:03}. The slit was 1$^{\prime\prime}$ wide, giving a resolution of 25,000. One 20-minute and two 30-minute exposures were taken, in addition to the usual Th-Ar and flatfield calibration frames. We use only the red side of the spectrograph, covering the wavelength range $9300-4900\ \mbox{\AA}$. The three exposures were combined to make the final spectrum for analysis. These data were then reduced using the procedure described in \citet{johnson:08}.

At its peak, OGLE-2007-BLG-514S was magnified by a factor of $\sim$1300 as it crossed a cusp in the caustic, brightening this $I\sim21$ mag star to approximately 13.3 mag. This is probably the faintest star for which a high-resolution spectrum has been obtained. Finite source effects, in particular the magnification of the limb relative to the center occur during such crossings. We have modeled the effect of differential limb magnification on the spectrum, and find that in more extreme cases than the actual observations discussed here, the change in the derived temperature is $<40$ K, in microturbulent velocity is $<0.08$ km s$^{-1}$, and in surface gravity is $<0.1$ dex, all much smaller than the uncertainties in our atmospheric parameters (see \S\ref{sec:error analysis}) due to the S/N of the spectrum, blending of lines, and other effects that appear in all spectra, whether magnified or not (Johnson et al.\ 2009, in prep). Johnson et al.\ find that ignoring differential limb magnification when analyzing the spectrum results in abundance errors $<0.05$ dex.

\begin{figure}
\plotone{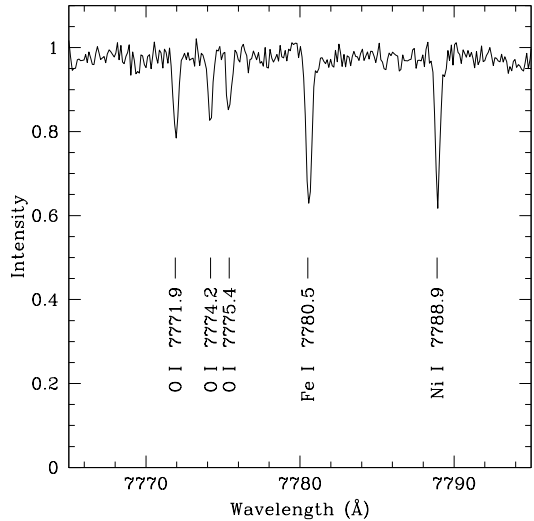}
\caption{Spectrum of OGLE-2007-BLG-514S in the region surrounding the O I triplet where the S/N$\sim 50$. The spectrum is shown shifted to the rest frame.}\label{fig:Otriplet}
\end{figure}

We measure the per pixel signal-to-noise ratio (S/N) in the continuum regions at three different wavelengths to provide a sense how the S/N deteriorates due to reddening. The spectrum has a S/N  of $\sim 50$ at $8000\ \mbox{\AA}$, $\sim 30$ at $7000\ \mbox{\AA}$, and $\sim 20$ at $6000\ \mbox{\AA}$. Bluer than that, the S/N additionally suffers from reduced flux from the star due to its relatively cool temperature and the strong metal-absorption lines in this portion of the spectrum. Figure \ref{fig:Otriplet} gives an example of the quality of the data in the red portion of the spectrum.

For OGLE-2007-BLG-514S, we measure an average radial velocity of $186.9 \pm 0.4$ km s$^{-1}$ from ten unblended iron lines with accurate laboratory wavelengths. At the time of observation, the heliocentric correction amounted to -28.08 km s$^{-1}$. Adding this correction, the heliocentric radial velocity of OGLE-2007-BLG-514S is $158.8 \pm 0.4$ km s$^{-1}$. The diversity of heliocentric radial velocities for all published bulge dwarfs in Figure \ref{fig:location} reflects the large velocity dispersion of stars in the bulge (e.g.\ \citealt{howard:08}).

\section{Abundance Analysis}
The linelist used for this analysis consists of a combination of the lists from \citet{johnson:08} and \citet{cohen:08}. Table \ref{table:lines} lists the transition probabilities and measured equivalent widths (EW) for lines in both OGLE-2007-BLG-514S and the Sun. We measure EW using SPECTRE\footnote{http://verdi.as.utexas.edu/spectre.html} for each line individually (C. Sneden, 2007, private communication).

\begin{deluxetable}{cccccc}
\tablewidth{0pc}
\tablecolumns{7}
\tablecaption{Line Parameters and Equivalent Widths}
\tablehead{\colhead{Ion} & \colhead{Wavelength} & \colhead{E.P.} & \colhead{$\log gf$} & \colhead{EW$_\mathrm{star}$} & \colhead{EW$_\odot$} \\
\colhead{} & \colhead{(\AA)} & \colhead{(eV)} & \colhead{} & \colhead{(m\AA)} & \colhead{(m\AA)}}
\startdata
Ba II & 6496.900 & 0.60 & -0.377 & syn & syn \\
Ca I & 5260.387 & 2.52 & -1.780 & 36.6 & 31.5 \\
Ca I & 5261.704 & 2.52 & -0.450 & 126.0 & 101.3 \\
Ca I & 5512.980 & 2.93 & -0.560 & 100.2 & 89.5 \\
Ca I & 5581.965 & 2.52 & -0.530 & 144.7 & 95.5 \\
Ca I & 5601.277 & 2.53 & -0.240 & 153.8 & 119.2 \\
Ca I & 5867.562 & 2.93 & -1.610 & 54.1 & 21.5 \\
Ca I & 6122.217 & 1.89 & -0.360 & 260.9 & 191.1 \\
Ca I & 6166.439 & 2.52 & -1.170 & 111.6 & 69.8 \\
Ca I & 6169.042 & 2.52 & -0.840 & 138.4 & 92.4 \\
$\vdots$ &$\vdots$ & $\vdots$ & $\vdots$ & $\vdots$ & $\vdots$ \\
\enddata\label{table:lines}
\tablecomments{Table 1 is published in its entirety in the electronic edition of \textit{The Astrophysical Journal}. A portion is shown here for guidance regarding its form and content.}
\end{deluxetable}

For elements with hyperfine splitting, namely Ba, Mn, V, and Sc, we use TurboSpectrum \citep{alvarez:98}, a 1-dimensional LTE code, to create a synthetic spectrum for the region surrounding the desired line and adjust the elemental abundance\footnote{We adopt the standard spectroscopic notation that $\mathrm{[X/Y]}\equiv\log_{10}(N_X/N_Y)_\bigstar-\log_{10}(N_X/N_Y)_\odot$ and $\log\epsilon(X)\equiv\log_{10}(N_X/N_H)+12.0$ for elements X and Y.} to match the observed spectrum. The hyperfine splitting A-values and, if available, B-values for the dominant isotope of vanadium, V-51, come from \citet{cochrane:98}, \citet{palmeri:95}, \citet{lefebvre:02}, and \citet{childs:79}. The HFS constants for the other elements are taken from the sources listed in \citet{johnson:06}.

\subsection{Atmospheric Parameters}\label{sec:atmospheric parameters}

We interpolated model atmospheres over a range in effective temperature, microturbulent velocity, surface gravity, and metallicity using a grid of the ATLAS9 models\footnote{http://wwwuser.oat.ts.astro.it/castelli/grids.html} with new opacity distribution functions \citep{castelli:03}. TurboSpectrum is used to derive elemental abundances from the measured EWs for each model atmosphere.

\begin{figure}
\plotone{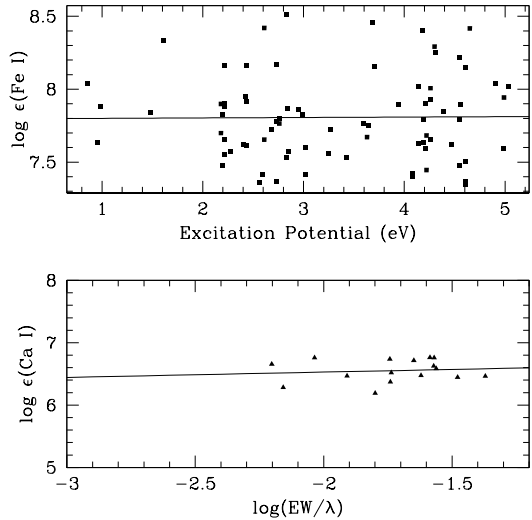}
\caption{\emph{Top:} Absolute Fe I abundance verses excitation potential demonstrating excitation equilibrium. \emph{Bottom:} Absolute Ca I abundance verses reduced EW. Solid lines give the weighted-least-squares fit to the abundances.}\label{fig:ca}
\end{figure}

Adhering to standard practice, the effective temperature (T$_{\mathrm{eff}}$) is found by ensuring excitation equilibrium for Fe I lines, pictorially represented by a zero slope in abundance with respect to excitation potential (EP). Microturbulent velocity ($\xi$) is determined by requiring a zero slope in the diagram of Ca I abundance verses reduced equivalent width ($\log(EW/\lambda)$). Both of these relationships are shown in Figure \ref{fig:ca}. The surface gravity ($\log \mathrm{g}$) is determined by requiring ionization equilibrium, i.e. that the average abundance of Fe I lines equals the average abundance from Fe II lines to within 0.02 dex. We then set the input model atmosphere metallicity to the [Fe I/H] determined relative to our measured solar value and iteratively adjust $\log \mathrm{g}$ and [Fe/H] until the metallicities match. To summarize our method, we determine the atmospheric model parameters $$m_j=(T_\mathrm{eff}, \xi, \log\mathrm{g}, \mathrm{[Fe/H]})$$ using the following observables $o_i$:
\begin{list}{}{}
\item $o_1$:  slope of $\log(\mathrm{Fe\ I})$ v.\ E.P.
\item $o_2$:  slope of $\log(\mathrm{Ca\ I})$ v.\ $\log(EW/\lambda)$.
\item $o_3$:  $\log\epsilon(\mathrm{Fe\ I})-\log\epsilon(\mathrm{Fe\ II})$ for the model.
\item $o_4$:  $\mathrm{[Fe\ I/H]}_\mathrm{model}-\mathrm{[Fe\ I/H]}_\mathrm{input}$.
\end{list}
where $m_j$ and $o_i$ are both four dimensional vectors. The best fit model $m_j^0$ gives T$_{\mathrm{eff}}=5600$ K, $\xi=1.5$ km s$^{-1}$, $\log \mathrm{g}=4.2$ dex, and $\mathrm{[Fe/H]}=0.33$ dex.

\subsection{Error Analysis}\label{sec:error analysis}

We present a rigorous error analysis that propagates the uncertainties in the observables to uncertainties in atmospheric parameters and elemental abundances. We begin by writing each observable as a linear combination of deviations from the best fit model:
\begin{equation}\label{eq:observables}
o_i = o_i^0 + \sum_{j=1}^4 b_{ij}(m_j - m_j^0)
\end{equation}
where $b_{ij}=\partial o_i/\partial m_j$. Varying one atmospheric parameter at a time, we create a series of models with $\Delta m_1=\pm 160$ K, $\Delta m_2=\pm 0.3$ km s$^{-1}$, $\Delta m_3=\pm 0.4$ dex, and $\Delta m_4=\pm 0.2$ dex. These $\Delta m_j$ are of order the size of the model atmosphere grid since the grid is approximately linear on those scales. Adjusting a model atmosphere by $\Delta m_j$ will change each of the observables by $(o_i-o_i^0)$. These partial derivatives $b_{ij}$ are given in Table \ref{table:derivatives}.

\begin{deluxetable}{c|cccc}
\tablewidth{0pc}
\tablecaption{Partial Derivatives of Observables with Respect to Atmospheric Model Parameters}
\tablehead{\begin{large}$b_{ij}$\end{large} & \begin{large}$\frac{\partial o_i}{\partial m_1}$\end{large} &
\begin{large}$\frac{\partial o_i}{\partial m_2}$\end{large} & \begin{large}$\frac{\partial o_i}{\partial m_3}$\end{large} &
\begin{large}$\frac{\partial o_i}{\partial m_4}$\end{large} \\[-12pt]}
\startdata
\begin{large}$\frac{\partial o_1}{\partial m_j}$\end{large} & \begin{large}$\frac{-0.03}{160}$\end{large} & \begin{large}$\frac{+0.02}{0.3}$\end{large} &
\begin{large}$\frac{+0.00}{0.4}$\end{large} &
\begin{large}$\frac{+0.00}{0.2}$\end{large} \\[8pt]
\begin{large}$\frac{\partial o_2}{\partial m_j}$\end{large} & \begin{large}$\frac{+0.04}{160}$\end{large} & \begin{large}$\frac{-0.06}{0.3}$\end{large} & \begin{large}$\frac{-0.23}{0.4}$\end{large} &
 \begin{large}$\frac{+0.07}{0.2}$\end{large} \\[8pt]
\begin{large}$\frac{\partial o_3}{\partial m_j}$\end{large} & \begin{large}$\frac{+0.20}{160}$\end{large} & \begin{large}$\frac{-0.04}{0.3}$\end{large} & \begin{large}$\frac{-0.22}{0.4}$\end{large} & \begin{large}$\frac{-0.04}{0.2}$\end{large} \\[8pt]
\begin{large}$\frac{\partial o_4}{\partial m_j}$\end{large} & \begin{large}$\frac{+0.11}{160}$\end{large} & \begin{large}$\frac{-0.10}{0.3}$\end{large} & \begin{large}$\frac{-0.04}{0.4}$\end{large} & \begin{large}$\frac{-0.17}{0.2}$\end{large} \\[-2pt]
\enddata\label{table:derivatives}
\end{deluxetable}

Since we have four observables, Equation (\ref{eq:observables}) produces a system of equations, which can be solved for the atmospheric model parameters. Inverting the matrix of $b_{ij}$ yields a matrix with elements $c_{ij}$.

Let $\sigma_i$ be the error in the measurement $o_i$. We treat the errors $\sigma_i$ as independent. Then the uncertainties in the model parameters are given by
\begin{equation}\label{eq:paramerr}
\sigma(m_i) = \sqrt{\sum_{k=1}^4 c_{ik}^2\sigma_k^2}.
\end{equation}

The errors $\sigma_1$ and $\sigma_2$ are the uncertainty in the slope of the linear least squares fit to $\log\epsilon(\mathrm{Fe\ I})$ v. E.P.\ and $\log\epsilon(\mathrm{Ca\ I})$ v.\ $\log(EW/\lambda)$, respectively. Uncertainties in EW and oscillator strengths produce a standard error of the mean abundance of 0.03 dex for Fe I and 0.10 dex for Fe II. We added in quadrature the standard error of the mean for Fe I and Fe II average abundances to find $\sigma_3$. We take the standard error of the mean Fe I abundance to be $\sigma_4$. Substituting these values into Equation (\ref{eq:paramerr}) yields the uncertainties in atmospheric parameters: $\sigma(T_\mathrm{eff})=180$ K, $\sigma(\xi)=0.5$ km s$^{-1}$, $\sigma(\log\mathrm{g})=0.3$ dex, and $\sigma(\mathrm{[Fe/H]})=0.15$ dex. These uncertainties, along with the values of the model atmosphere parameters, are given in Table \ref{table:stellarparam}.

\begin{deluxetable}{ccccc}
\tablewidth{0pc}
\tablecaption{Stellar Parameters and Uncertainties for OGLE-2007-BLG-514S}
\tablehead{&\colhead{$T_\mathrm{eff}$} & \colhead{$\xi$} & \colhead{$\log\mathrm{g}$} & \colhead{[Fe/H]} \\ &\colhead{(K)} & \colhead{(km s$^{-1}$)} & \colhead{(dex)} & \colhead{(dex)}}
\startdata
$m_j$ & $5600$ & $1.5$ & $ 4.2$ & $0.33$ \\
$\sigma(m_j)$ & $180$ & $0.5$ & $ 0.3$ & $0.15$ \\
\enddata\label{table:stellarparam}
\end{deluxetable}

The next step is to find the uncertainty in the abundance of some element $X$. As before, we can write our measured quantity, in this case $X$, as
\begin{equation}
X = X_0+\sum_{j=1}^4 \kappa_j(m_j - m_j^0)
= X_0+\sum_{j=1}^4 \alpha_j(o_j - o_j^0)
\end{equation}
where $\kappa_j=\partial X/\partial m_j$ and
\begin{equation}
\alpha_j\equiv \sum_{k=1}^4 \kappa_k c_{kj}.
\end{equation}
We found $\kappa_j$ by running the series of models with $\Delta m_j$ through TurboSpectrum or, for elements with hyperfine splitting, by creating a synthetic spectrum for each model. Thus, the error in $\log\epsilon(X)$ is
\begin{equation}\label{eq:x}
\sigma(X) = \sqrt{\sigma^2(X_0) + \sum_{k=1}^4 \alpha_k^2\sigma_k^2}
\end{equation}
where $\sigma(X_0)$ is the standard error of the mean abundance of element $X$ at fixed stellar model parameters. For elements with either one measured line or a synthesized spectrum, we take the change in abundance required to compensate for a generous offset in the continuum placement as the error in measuring $X_0$. Otherwise, we use the rms in abundance for elements with many measured lines.

Often, however, we are interested in the ratio of two elements X and Y, written [X/Y]. The uncertainty in this quantity is
\begin{equation}\label{eq:xfe}
\sigma([X/Y])=\sqrt{\sigma(X)^2+\sigma(Y)^2-2\sum_{k=1}^4 \alpha^X_k \alpha^Y_k\sigma_k^2}
\end{equation}
The uncertainties calculated with Equations (\ref{eq:x}) and (\ref{eq:xfe}) are given in Table \ref{table:abundances}.

\subsection{Photometric Parameters}\label{sec:photometric parameters}

De-reddened colors and magnitudes of the microlensed source are estimated using standard microlensing techniques \citep{yoo:04}. We assume that the reddening to the microlensed source is the same as the reddening to the red clump giants in the bulge. \citet{stanek:00}, \citet{sumi:04}, and \citet{bennett:08} report a de-reddened $(V-I)_0$ for the bulge red clump in the range 1.028-1.066. Taking $(V-I)_{RC} = 1.05$, photometric observations from $\mu$FUN SMARTS CTIO 1.3m telescope determined the dereddened Johnson-Cousins color and magnitude of OGLE-2007-BLG-514S are $(V-I)_0 = 0.70 \pm 0.05$ and $I_0 = 19.2 \pm 0.1$ mag, respectively. The similarity between OGLE-2007-BLG-514S's color and the Sun's $(V-I)_\odot = 0.688 \pm 0.014$ \citep{holmberg:06} implies that the temperatures of these stars must also be close. Using the relationship between temperature, color, and metallicity in \citet{ramirez:05} (updated from the color-temperature relation in \citealt{alonso:96}), we estimate a photometric temperature of $5760 \pm 200$ K. This photometric temperature agrees with the spectroscopic temperature within error.

\subsection{Solar Spectrum}\label{sec:solar}

\begin{deluxetable*}{lccccrcc}
\tablecolumns{8}
\tablewidth{0pc}
\tablehead{\colhead{Ion} & \colhead{$\log\epsilon(X)$} & \colhead{$\sigma(X)$} & \colhead{[X/Fe I]\tablenotemark{a}} & \colhead{$\sigma(\mathrm{[X/Fe\ I]})$\tablenotemark{a}} & \colhead{$N_\mathrm{lines}$} & \colhead{Solar} & \colhead{Solar}\\ & & & & & & \colhead{Measured} & \colhead{GS98}}
\startdata
O  I & 9.08 & 0.21 & -0.06 & 0.22 & 3 & 8.80 & 8.83 \\
Na I & 6.86 & 0.10 & 0.27 & 0.12 & 5 & 6.26 & 6.33 \\
Mg I & 8.05 & 0.08 & -0.01 & 0.14 & 12 & 7.73 & 7.58 \\
Si I & 8.01 & 0.11 & 0.11 & 0.11 & 20 & 7.56 & 7.55 \\
K  I & 5.41 & 0.27 & -0.16 & 0.25 & 1 & 5.24 & 5.12 \\
Ca I & 6.55 & 0.12 & -0.07 & 0.07 & 16 & 6.29 & 6.36 \\
Sc II\tablenotemark{b}  & 3.30 & 0.32 & -0.23 & 0.25 & 2 & 3.20 & 3.17 \\
Ti I & 5.56 & 0.26 & 0.37 & 0.25 & 6 & 4.86 & 5.02 \\
Ti II & 5.58 & 0.55 & 0.49 & 0.53 & 2 & 4.76 & 5.02 \\
V  I\tablenotemark{b}  & 4.30 & 0.26 & 0.17 & 0.29 & 5 & 3.80 & 4.00 \\
Cr I & 6.22 & 0.29 & -0.03 & 0.28 & 8 & 5.91 & 5.67 \\
Cr II & 6.27 & 0.38 & 0.30 & 0.35 & 3 & 5.64 & 5.67 \\
Mn I\tablenotemark{b}  & 5.80 & 0.24 & 0.17 & 0.23 & 1 & 5.30 & 5.39 \\
Fe I & 7.81 & 0.15 & 0.33 & 0.15 & 80 & 7.47 & 7.50 \\
Fe II & 7.79 & 0.21 & -0.08 & 0.15 & 7 & 7.54 & 7.50 \\
Co I & 5.32 & 0.17 & 0.07 & 0.18 & 4 & 4.91 & 4.92 \\
Ni I & 6.72 & 0.15 & 0.15 & 0.07 & 30 & 6.24 & 6.25 \\
Ba II\footnotemark{b}  & 2.00 & 0.39 & -0.63 & 0.29 & 1 & 2.30 & 2.13 \\
\enddata\label{table:abundances}
\tablenotetext{a}{[X/Fe I] and $\sigma(\mathrm{[X/Fe\ I]})$ are given for all species except Fe I, where [Fe I/H] and $\sigma(\mathrm{[Fe\ I/H]})$ are given, respectively.}
\tablenotetext{b}{Abundance determined by creating a synthetic spectrum.}
\end{deluxetable*}

We performed an identical analysis using the same set of lines, line parameters, and model atmosphere grid as for OGLE-2007-BLG-514S on a high-resolution solar spectrum also taken by the MIKE spectrograph of reflected light from Jupiter's moon Ganymede \citep{bensby209:09}. We fixed the well-known model solar atmosphere parameters at $T_\mathrm{eff}=5777$ K, $\log\mathrm{g}=4.4$ dex, and $\mathrm{[Fe/H]}=0$ dex and adjusted $\xi$ until excitation equilibrium was achieved at $\xi=0.88$ km s$^{-1}$. Table \ref{table:abundances} presents both the solar abundances from our measurements and the literature values from \citet{grevsauv}. All final elemental abundances given in Table \ref{table:abundances} were normalized to the measured solar spectrum. This differential analysis is appropriate for bulge dwarfs since they are similar type stars to the Sun and the same lines are visible in both stars.

For the solar spectrum, we find $\log\epsilon(\textrm{Ba})_\odot = 2.3$ dex from the Ba II line at 6496.898 \AA, which is the only Ba line we measured in the OGLE-2007-BLG-514S spectrum. Our measurement is higher than both the photospheric abundance of $2.13 \pm 0.05$ dex and meteoritic abundance of $2.22 \pm 0.02$ dex from \citet{grevsauv}\footnote{We compare to \citet{grevsauv} since they also use a 1-dimensional model atmosphere. Recent 3-D models yield a higher photospheric Ba abundance of $2.18 \pm 0.09$ dex \citep{aspulund:09}.}. The measurement error for Ba in our very high S/N solar spectrum is negligible compared to possible systematic effects. The disagreement in literature meteoritic and photospheric Ba abundances indicates that resolving $\sim0.09$ dex differences is an ongoing issue for Ba measurements. The photospheric absolute Ba abundance depends sensitively on the specifics of the solar model atmosphere and the line parameters. In particular, we find that changing the model atmosphere parameter $\xi$ by $+0.3$ km s$^{-1}$ results in a $+0.2$ dex change in the Ba abundance. The absolute Ba abundance is also influenced by the van der Waals damping coefficient \citep{mashonkina:06}; although we used the most up-to-date values, they may not be absolutely correct. Fortunately, uncertainties in the absolute abundance due to errors in the model atmosphere and line parameters cancel to first order when considering the stellar abundance \textit{relative} to the solar values found using the same technique. For this reason, we use the metallicity normalized to a solar spectrum consistent with the stellar analysis if possible when comparing OGLE-2007-BLG-514S with other stars.

\section{Bulge Membership}

From OGLE-2007-BLG-514S's high heliocentric radial velocity ($158.8 \pm 0.4$ km s $^{-1}$), we infer that the source is in the bulge rather than the disk. Disk stars at the same galactic longitude have radial velocities
$$\mathrm{RV}_\mathrm{disk} \sim \frac{220\mathrm{\ km\ s}^{-1} R_0 \sin l}{\eta} \sim 80\mathrm{\ km\ s}^{-1}\left( \frac{1 \mathrm{\ kpc}}{\eta}\right) $$
where $R_0=8$ kpc and $\eta$ is the distance from the Galactic Center to OGLE-2007-BLG-514S. We limit $\eta > 1 \mathrm{\ kpc}$ to exclude the bulge. Given the disk's radial velocity dispersion of $\sigma\sim 35$ km s$^{-1}$ \citep{dehnen:98}, the probability is low that a disk star would achieve a radial velocity as high as OGLE-2007-BLG-514S. By contrast, OGLE-2007-BLG-514S's radial velocity is typical of the bulge which has a mean radial velocity of $-5 \pm 14$ km s$^{-1}$ with a dispersion of $\sigma = 112 \pm 10$ km s$^{-1}$ \citep{howard:08}.

We would naively expect OGLE-2007-BLG-514S to lie in the bulge because at $(l, b) = (2.62, -1.63)$ the density of source stars along the line of sight is sharply peaked within $R_0 \sin b \sim 200$ pc of the Galactic Center, where $R_0 = 8$ kpc \citep{gould:00}. Additionally, the microlensing-derived color and dereddened apparent $I$-band magnitude relative to the bulge red clump stars are consistent with the source being a dwarf star with the same reddening and distance as the red clump.

\begin{figure}
\includegraphics[width=3.3in,clip=true,trim=.0in .0in 2.8in 0in]{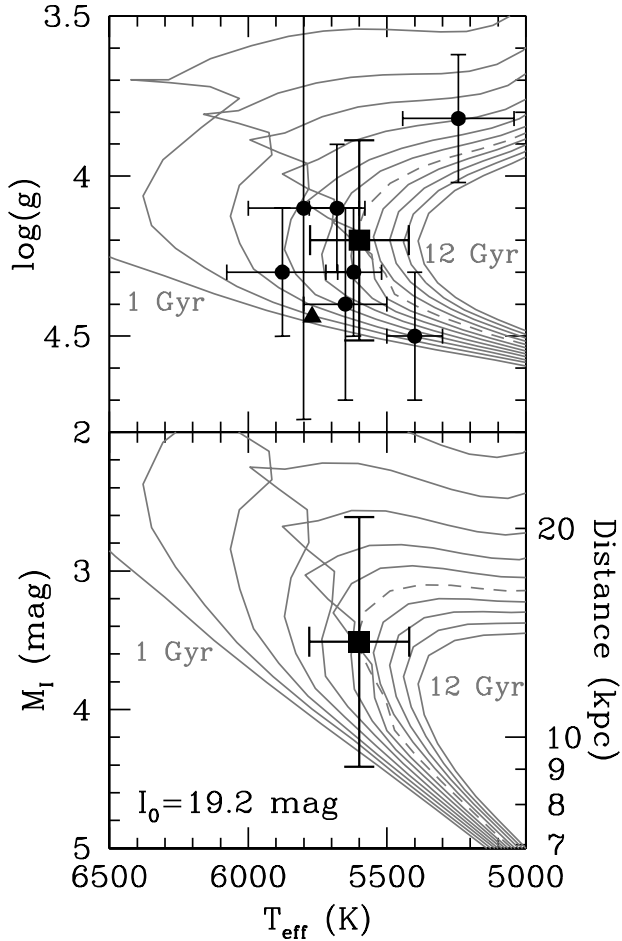}
\caption{Y$^2$ isochrones (gray) are shown for [Fe/H]=0.33 ($\mathrm{Z}=0.0359$) with no alpha enhancement for ages 1--12 Gyr in increments of 1 Gyr. The dashed curve shows the best-fit isochrone (8 Gyr). \emph{Top:} The positions of OGLE-2007-BLG-514S (square), the Sun (triangle), and the other bulge dwarfs and subgiant (circles) on the H-R diagram. \emph{Bottom:} Absolute magnitudes and the corresponding distance assuming a dereddened apparent magnitude $I_0=19.2$ mag of OGLE-2007-BLG-514S.}
\label{fig:isochrones}
\end{figure}

Using our spectroscopic measurements of T$_{\mathrm{eff}}$ and $\log \mathrm{g}$, we can estimate OGLE-2007-BLG-514S's distance modulus (DM). In the top panel of Figure \ref{fig:isochrones}, we compare the position of OGLE-2007-BLG-514S on the Hertzsprung-Russell diagram with the other bulge dwarfs and theoretical isochrones. We employ the Yonsei-Yale (Y$^2$) isochrones \citep{yi:01,demarque:04} with $\mathrm{[Fe/H]}=0.33$ ($\mathrm{Z}=0.0359$) and $[\alpha/\mathrm{Fe}]=0$ to match OGLE-2007-BLG-514S. Figure \ref{fig:isochrones} shows the other bulge dwarfs and the Sun for comparison. Since isochrones depend on a star's metal abundance, the displayed isochrones are only relevant for stars with the same metallicity as OGLE-2007-BLG-514S.

In the bottom panel of Figure \ref{fig:isochrones}, we determine where a star with $\log\mathrm{g}$ and best fit age (8 Gyr) falls on the M$_\mathrm{I}-T_\mathrm{eff}$ plane. Transferring the errors in $\log\mathrm{g}$ to the M$_\mathrm{I}-T_\mathrm{eff}$ plane at constant $T_\mathrm{eff}$ translates to a 0.9 mag uncertainty in $M_I$. Combining OGLE-2007-BLG-514S's dereddened apparent $I$-band magnitude ($I_0=19.2\pm 0.1$ mag) with an absolute $\mathrm{M}_\mathrm{I}=3.5 \pm 0.9$ mag from the isochrones yields $\mathrm{DM}= 15.7 \pm 0.9$ mag. Assuming the bulge extends 1 kpc from the Galactic Center, it encompasses the range $14.25 \leq  \mathrm{DM}_\mathrm{bulge} \leq 14.74$ mag along the line of sight. The DM to OGLE-2007-BLG-514S is in $1.3\sigma$ conflict with the Galactic Center, but is consistent with the far side of the bulge. Including the uncertainties in $T_\mathrm{eff}$ and [Fe/H] will slightly increase the uncertainty in the DM. For example, changing the [Fe/H] of the isochrones by our $+0.15$ dex error in metallicity resulted in a $-0.09$ mag difference in $M_I$.

From the finite-source effects, the relative lens-source proper motion of OGLE-2007-BLG-514 is precisely determined to be $3.0\ \rm mas\ yr^{-1}$. While the proper motion is typical of bulge-bulge events, it is also compatible with the lensing star lying in the bulge and source in the far disk. This does not help us discriminate between the far disk or bulge senarios.

\section{The Metallicity Distribution}\label{sec:MDF}
\begin{figure}
\plotone{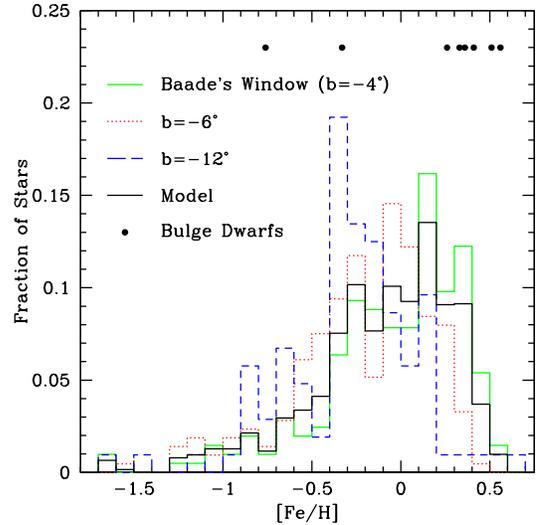}
\caption{Normalized iron metallicity distribution function (MDF) of stars in the bulge. The metalicity of giants is measured along the minor axis of the bulge at Galactic latitudes $b=-4^\circ$ (Baade's window), $-6^\circ$, and $-12^\circ$ by \citet{zoccali:08}. The black histogram shows our model bulge giant MDF that accounts for the different radial distribution of bulge dwarfs from the GC as described in \S\ref{sec:MDF}. The eight bulge dwarfs are shown in a linear distribution above the histograms.}
\label{fig:mdf}
\end{figure}

Three-quarters of the published bulge dwarfs exhibit high metallicities; only the sub-giant OGLE-2008-BLG-209S and OGLE-2009-BLG-076S have sub-solar [Fe/H]. Figure \ref{fig:mdf} compares the bulge dwarf MDF to that of giants located in the three fields marked in Figure \ref{fig:location} \citep{zoccali:08}. In contrast to the concentration of bulge dwarfs at high metallicity ($<[\mathrm{Fe/H}]>=0.17$ dex), the bulge giants have a broad, asymmetric MDF centered just below solar metallicity which extends to $[\mathrm{Fe/H}]\sim-1.5$ dex. From the bulge dwarf data published up to this point, it appears that the profile of the bulge dwarf MDF differs from that of the bulge giants.

Figure \ref{fig:mdf} demonstrates that the shape of the giant MDF depends on position in the bulge. The bulge dwarfs are found at a variety of apparent angular separations from the Galactic Center, with only the subgiant falling in a field where the giant MDF is known. To make a meaningful comparison between bulge dwarfs and giants, we must account for how the observed bulge giant MDF would change if the giants had the same spatial distribution as the dwarfs.

The bulge giant MDF is only measured in a handful of low-reddening windows. \citet{zoccali:08} find the MDF of giants at three points along the minor axis of the bulge and detect a small radial gradient between Baade's window ($b=-4^\circ$) and $b=-12^\circ$. Between $(l,b)=(0^\circ,-1^\circ)$ and Baade's window, \citet{rich:07} find no evidence of an iron abundance gradient. Assuming the trend along the minor axis applies to the entire bulge, we adopt a simple model that consists of a constant MDF inside Baade's window with a small radial gradient beyond it. The dashed circles in Figure \ref{fig:location} show the radial distribution of the bulge dwarfs relative to the three fields of giants. The location of each dwarf is then associated with a model giant MDF. For a dwarf positioned between two Zoccali et al. fields, the corresponding model giant MDF is the average of the MDFs in those two fields weighted linearly by the radial proximity of the star and field. If a dwarf lies inside Baade's window, the model MDF is the same as the MDF in Baade's window. These eight model MDFs are averaged to find the total MDF pictured in Figure \ref{fig:mdf}.

\begin{figure}
\plotone{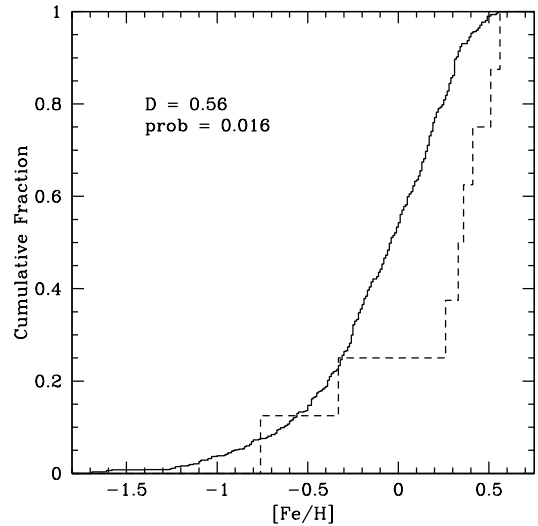}
\caption{Two-sample Kolmogorov-Smirnov test between the bulge dwarf MDF (dashed) and the model bulge giant MDF (solid).}
\label{fig:KStest}
\end{figure}

To determine whether the giant and dwarf MDFs share the same underlying distribution, we perform a two-sided Kolmogorov-Smirnov (K-S) test on the most up-to-date sample of eight bulge dwarfs and the model giant MDF. Figure \ref{fig:KStest} gives the cumulative fraction of stars with a lower $[\mathrm{Fe/H}]$ in both populations. The maximum vertical distance between the cumulative distribution of dwarfs and giants is $D = 0.56$. Calculating a probability from the D statistic requires the number of stars in each distribution. For the model, we took the equivalent number of stars to be the average of the number of stars in each field weighted in the same fashion as for the model MDF. This yields a probability of 1.6\% that the dwarfs and giants have been drawn from the same distribution.

\begin{deluxetable*}{lccccccccccccccc}
\tabletypesize{\scriptsize}
\tablewidth{0pc}
\tablecaption{Literature Sources}
\tablehead{\colhead{Source} & \colhead{O} & \colhead{Na} & \colhead{Mg} & \colhead{Si} & \colhead{K} & \colhead{Ca} & \colhead{Sc} & \colhead{Ti} & \colhead{V} & \colhead{Cr} & \colhead{Mn} & \colhead{Co} & \colhead{Ni} & \colhead{Ba} & \colhead{Y}}
\startdata
\sidehead{\textit{Bulge:}}
\citet{fulbright:07} & X & X & X & X &  & X &  & X &  &  &  &  &  &  &  \\
\citet{ryde:09} & X &  &  & X &  &  &  & X &  & X &  &  & X &  &  \\
\citet{rich:05} & X &  & X & X &  & X &  & X &  &  &  &  &  &  &  \\
\citet{rich:07} & X &  & X & X &  & X &  & X &  &  &  &  &  &  &  \\
\citet{lecureur:07} & X & X & X &  &  &  &  &  &  &  &  &  &  &  &  \\
\citet{cunha:06} & X & X &  &  &  &  &  & X &  &  &  &  &  &  &  \\
\sidehead{\textit{Halo and Disk:}}
\citet{reddy:06} & X & X & X & X &  & X & X & X & X & X & X & X & X & X & X \\
\citet{reddy:03} & X & X & X & X & X & X & X & X & X & X & X & X & X & X &  \\
\citet{feltzing:07} &  &  &  &  &  &  &  &  &  &  & X &  &  &  &  \\
\citet{bensby:05} &  &  & X & X &  & X &  & X &  & X &  &  & X & X & X \\
\citet{bensby:04} & X &  &  &  &  &  &  &  &  &  &  &  &  &  &  \\
\citet{bensby:03} &  & X & X &  &  &  &  &  &  &  &  &  &  &  &  \\
\citet{chen:03} & X & X & X & X & X & X & X & X & X & X & X & X & X & X &  \\
\citet{zhang:06} &  &  &  &  & X &  &  &  &  &  &  &  &  &  &  \\
\citet{mashonkina:07} &  &  &  &  &  &  &  &  &  &  &  &  &  & X & X \\
\enddata\label{table:litsources}
\end{deluxetable*}

If the underlying distribution for dwarfs and giants is actually identical, this low probability could could stem from either a systematic error in measuring the [Fe/H] in dwarfs or giants \citep{cohen:09} or from small number statistics. Addressing this last concern, \citet{bensbyIAU:09} presents preliminary [Fe/H] measurements for six more bulge dwarfs. They fill the gap in the bulge dwarf MDF at intermediate metallicity, which would weaken the K-S test if their preliminary [Fe/H] are confirmed. However, two additional dwarfs observed at the end of the bulge season have supersolar preliminary [Fe/H]. Interpretation of the MDF is in a state of flux as the sample size rapidly expands.

The data published to date yield a 2$\sigma$ result in favor of an intrinsically different dwarf and giant MDF. Figure \ref{fig:mdf} suggests that metal-rich bulge stars do not become giants. One explanation is that some metal-rich stars lose enough mass through winds or some other mechanism on their ascent up the RGB that they do not make it to the horizontal branch and instead become He WDs \citep{cohen:08}. Although He WDs are traditionally thought to come from interacting binary systems, extreme mass loss on the RGB could produce He WDs from single stars (e.g.\ \citealt{castellani:93,dcruz:96}). Observational evidence for this idea first surfaced in a Galactic open cluster. \citet{hansen:05} invoked this mechanism to explain a brighter than expected peak in the luminosity function of the WD cooling sequence in the old metal-rich cluster NGC 6791 discovered by \citet{bedin:05}. In a followup spectroscopic study, \citet{kalirai:07} identified low-mass WDs, noted that chemically enriched stars lose a larger fraction of their mass, and suggested that that the high metallicity of the cluster causes mass loss. For field stars in the local Galactic disk, \citet{kilic:07} found evidence of this phenomenon through a population analysis of metal-rich stars and single low mass WDs.

Further scrutiny led some to dispute the idea that mass loss on the RGB produces He WDs in NGC 6791. \citet{vanloon:08} found little circumstellar dust around RGB stars in NGC 6791. Furthermore, their optical and infrared luminosity functions along the RGB agreed with expectations. However, the interpretation of the van Loon et al.\ results is complicated by the unknown properties of mass loss, leaving still open the question of the origin of He WDs in NGC 6791. As an alternative explanation, \citet{bedin:08} claim that the bright peak in the luminosity function arises naturally if $\sim$34\% of the WDs are actually unresolved binary WD+WD systems. Bedin et al.\ recalculate NGC 6791's binary fraction to fall in the range $>25\%-30\%$, a marked increase from the 14\% binary fraction determined by \citet{janes:97}.

The idea of mass loss on the RGB faces challenges not only in NGC 6791, but also in the bulge. If bulge stars peel off the RGB to become He WDs, the upper RGB luminosity function should thin out, but this is not observed in the fields studied by \citet{zoccali:03}.

A difference between the fraction of metal-rich dwarfs and giants appears in the local neighborhood as well as the bulge MDF. \citet{luck:07} find their sample of dwarfs within 15 pc of the Sun has a more pronounced high-metallicity tail than the giants do (see their Figure 9). Luck et al.\ attribute the dwarfs' metal enhancement to the high fraction of stars hosting close-in giant planets within their sample. The nature of microlensing events provides an unbiased sample of bulge dwarfs with respect to the presence of planets. If the dissimilarity between the bulge dwarf and giant MDFs is real, perhaps the mechanism responsible for it contributes to the difference seen in the local neighborhood.

\section{Chemical Evolution}

OGLE-2007-BLG-514S is a metal-rich ([Fe/H]$=0.33\pm0.15$) dwarf formed at the end of the star formation period in the bulge. The bulge's unique star formation history is reflected in its abundance patterns when compared to the disk and halo. We compare the OGLE-2007-BLG-514S's abundance measurements to six bulge dwarfs since detailed chemical abundances are not yet available for OGLE-2009-BLG-076S. Abundance trends for the comparison sample of bulge giants and disk and halo dwarfs are taken from the literature sources specified in Table \ref{table:litsources}.

\begin{figure*}
\centering
\includegraphics[width=6.5in,clip=true,trim=.0in .0in 1.7in 0in]{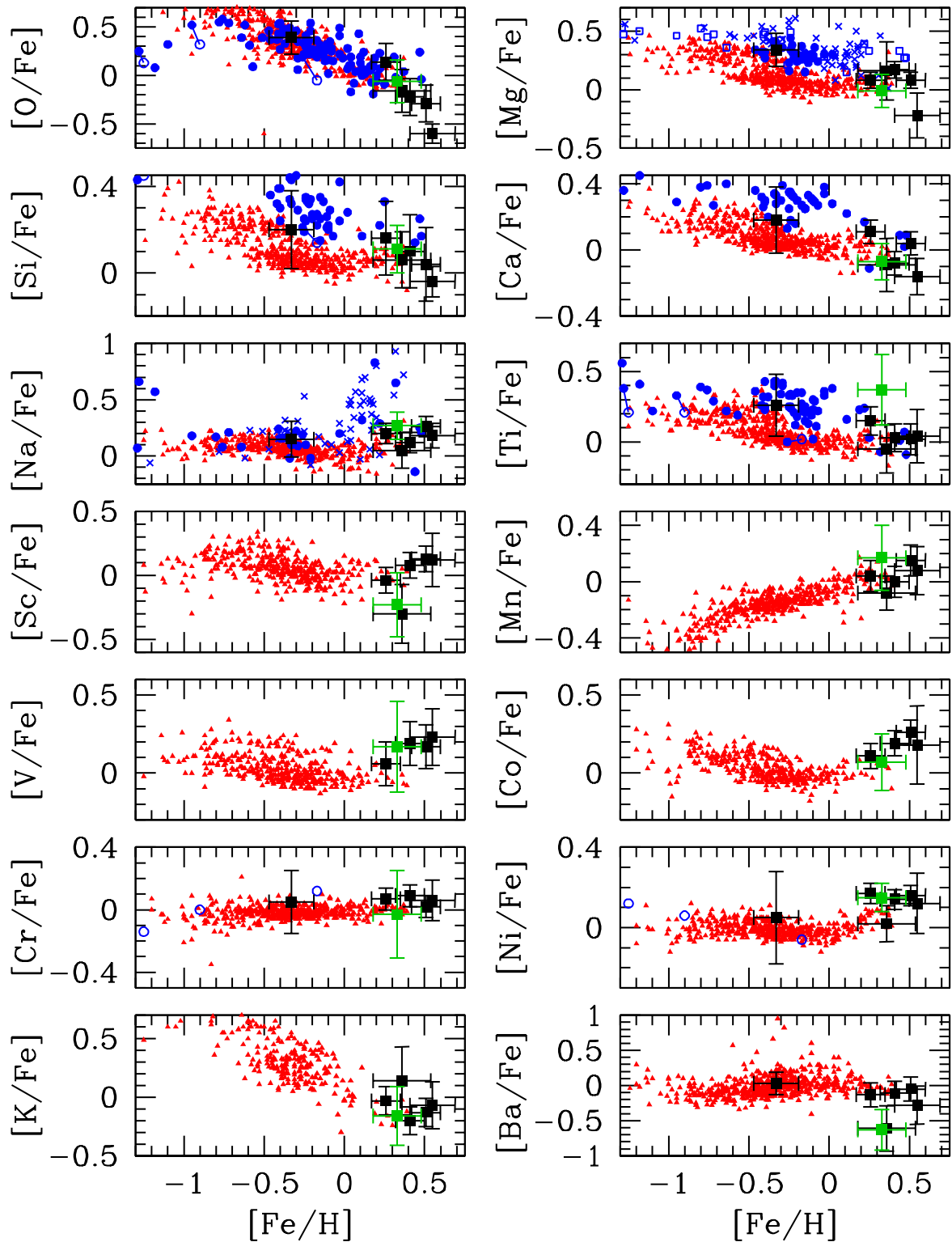}
\caption{Element abundance for bulge dwarfs (solid black squares) compared to bulge giants (blue) and field stars (open red triangles). Abundances for OGLE-2007-BLG-514S are calculated relative to our measured solar abundances (green). In specific cases, we indicate the literature source of bulge giant data. \citet{lecureur:07} is marked by blue crosses in the [Mg/Fe] and [Na/Fe] panels. The open squares in [Mg/Fe] are from \citet{fulbright:07}. \citet{ryde:09} remeasured abundances in selected stars from the Fulbright et al.\ sample. The open blue circles giving Ryde et al.'s measurement are connected to the corresponding Fulbright et al.\ point. Detailed chemical abundances are not yet available for OGLE-2009-BLG-076S.}
\label{fig:all}
\end{figure*}

\subsection{The $\alpha$-elements}\label{sec:alpha}

Alpha-elements (O, Mg, Si, Ca, and Ti) are primarily formed by massive stars and ejected in SNcc explosions, but iron is produced in both SNIa and SNcc. Stars forming from gas polluted by SNcc will have an alpha enrichment compared to those forming after delayed SNIa events begin to contribute a significant amount of iron to the ISM. Figure \ref{fig:all} shows that SNcc dominated the iron input into the bulge to a higher metallicity than in the disk. The OGLE-2007-BLG-514S is consistent with a solar $[\alpha/\mathrm{Fe}]$ indicating that it formed from metal enhanced gas toward the end of the bulge's star formation period.

Both \citet{lecureur:07} and \citet{fulbright:07} found an elevated [Mg/Fe] ratio that is nearly flat with [Fe/H]. OGLE-2007-BLG-514S attained a solar level of [Mg/Fe]$=-0.01\pm0.19$ dex, which is consistent with scatter in bulge stars at high metallicity.

\subsection{Sodium}

Below solar [Fe/H], the Na abundances of disk dwarfs and bulge giants are roughly flat and have similar dispersions. However, at $\mathrm{[Fe/H]}\gtrsim0$, the Na abundance of bulge giants is elevated relative to disk dwarfs and the dispersion increases \citep{mcwilliam:94,cunha:06,lecureur:07}. In the following discussion, we focus on the \citet{lecureur:07} analysis because their sample contributed most of the Na measurements for bulge giants with supersolar [Fe/H]. Since they were unable to explain the large dispersion with observational errors or uncertainties in the model atmospheres, Lecureur et al.\ investigated whether enhanced extra mixing polluted the Na surface abundance of the giants in their sample. Some globular clusters, like M4, display an O-Na anticorrelation thought to come from enhanced extra mixing on the upper RGB that penetrates deep enough in the H-burning shell to dredge up material enriched in N and Na and deficient in C and O (e.g.\ \citealt{kraft:94,denissenkov:06}). Enhanced extra mixing in M4 results in a factor of $\sim 2-3$ difference in C+N between AGB and RGB stars \citep{ivans:99,smith:05}. Although Lecureur et al.\ find an O-Na anticorrelation, they see no separation between red clump and RGB stars in the flat trend of [C+N/Fe] with metallicity (see Figure 8 in \citealt{lecureur:07}). They attribute the anticorrelation to [O/Fe] evolution and argue that the coincidence of bulge RGB and red clump stars in [C+N/Fe] demonstrates that internal mixing on the RBG does not reach below the CN-cycled layer. If true, the Na surface abundance in bulge giants should reflect the composition of the ISM at formation and match the Na abundance distribution of unmixed dwarf stars.

However, we do not observe the anticipated agreement. The bulge dwarfs appear more tightly clustered in [Na/Fe] than the widely scattered bulge giants do at [Fe/H$]\gtrsim0$. Furthermore, none of the bulge dwarfs observed so far exhibit the extreme [Na/Fe] ratios measured by \citet{lecureur:07} and instead are consistent with the disk dwarf trend.

This discrepancy between bulge dwarfs and giants has four possible explanations. First, a real difference exists between the bulge Na abundance at solar metallicity and at high metallicity where the giants and dwarfs are respectively concentrated. Unfortunately, we cannot test yet whether the extreme [Na/Fe] ratios in the bulge are confined to $0.0\lesssim\mathrm{[Fe/H]}\lesssim0.3$ since we lack published bulge dwarf abundances in that metallicity range. A second option is that internal mixing delves deeper than expected and the giants' surface abundances are corrupted by dredge up. Otherwise, either the giant measurements or the dwarf measurements are in error. We consider the last prospect unlikely. A conspiracy of errors would be required to produce the observed tight bulge dwarf relationship if the real bulge dwarf [Na/Fe] distribution matched the giants' high mean and large scatter. Additionally, the bulge dwarf [Na/Fe] measurements are a robust relative comparison to the Sun since they are main sequence G-stars.

\subsection{Iron-Peak Elements}

Although part of the iron-peak, Ti behaves like an alpha element with an enhanced abundance relative to Fe until [Fe/H]$\sim$0 and downward slope toward the solar ratio thereafter \citep{mcwilliam:94}. Figure \ref{fig:all} shows that [Ti/Fe]=0.37$\pm$0.25 dex for OGLE-2007-BLG-514S is a little high for its metallicity, but consistent with this trend within the (large) uncertainty.

The Fe-peak elements Sc, V, Cr, Mn, Co, and Ni are displayed in Figure \ref{fig:all}. The [V/Fe] ratio reaches higher values in the bulge than in the disk for metal rich stars. Since the bulge dwarfs have higher metallicity than are measured for disk dwarfs, we compare the average [V/Fe] of the five bulge dwarfs for which we have V information with the average the five most metal rich disk dwarfs. In the bulge, $<\mathrm{[V/Fe]}>=0.16\pm0.06$ dex while in the disk $<\mathrm{[V/Fe]}>=-0.04\pm0.03$ dex, where the uncertainties are rms deviations in the abundance measurements. Apparently, [V/Fe] increases at high metallicity in the bulge. More data are required to determine whether the disk displays a similar increase at comparatively high metallicity.

Otherwise the abundance measurements of OGLE-2007-BLG-514S and the other bulge dwarfs follow the disk trends within the errorbars. In particular, OGLE-2007-BLG-514S conforms to the observed trend in [Sc/Fe] and [Ni/Fe], which tend to be near solar for the halo, hin and thick disks, and the bulge.

\citet{mcwilliam:04} note that the Mn behaves similarly in the bulge and disk: it rises toward solar with increasing [Fe/H], symmetric to the $\alpha$-element's decline. \citet{gratton:89} suggested that the same mechanism can explain both trends and inferred that SNIa produce more Mn than SNcc. If SNIa did dominate the Mn production, $\alpha$-enhanced bulge stars should have low Mn. Further investigation by \citet{feltzing:07} revealed that [Mn/O] begins rising before SNIa contributions become important around [O/H$]\sim0$. This favors metallicity-dependent yields of SNcc over SNIa as the primary source of Mn. For OGLE-2007-BLG-514S, the solar [$\alpha$/Fe] ratios argue that both SNIa and SNcc have contributed and we cannot distinguish which type or combination of SN produced this star's near solar Mn abundance.

\subsection{Potassium}

We measured the K 7698 \AA\ resonance line in OGLE-2007-BLG-514S and the Sun. Since only one line was available for study in this star, we based the uncertainty on the changes in EW for different continuum placement. Although this line is affected by non-LTE effects, we did not apply a NLTE correction since they should be the same for the Sun and cancel out in the differential analysis.

Literature measurements show an increase in [K/H] with decreasing metallicity in the disk \citep{reddy:03,chen:03,zhang:06}. In Figure \ref{fig:all}, the K abundance of OGLE-2007-BLG-514S obeys the same trend. \citet{zhang:06} found a nearly constant [K/Mg] ratio ($<\mathrm{[K/Mg]}>=-0.08\pm0.01$ dex with 0.07 dex scatter) and inferred that K production is coupled to the $\alpha$-elements. For OGLE-2007-BLG-514S, we measure $\mathrm{[K/Mg]}=-0.15\pm 0.26$ consistent with this flat trend, despite the different chemical evolution history of the bulge and disk.

\subsection{Barium}

Figure \ref{fig:all} shows that Ba is underproduced relative to the Galactic disk in OGLE-2007-BLG-514S and MOA-2006-BLG-099S. Ba is predominantly generated in the $s$-process in the solar system ($\sim80\%$ according to \citealt{arlandini:99} models). OGLE-2007-BLG-514S's low [Ba/Fe] ratio could be explained if nucleosynthesis is dominated by the \textit{r}-process, which is a much less efficient Ba producer. Halo stars exhibit low [Ba/Fe] ratios since the \textit{r}-process was the only available channel for producing heavy elements in the early universe. If the \textit{r}-process dominates the bulge, then \textit{r}-process elements, like Eu, should show an enhancement. Although Eu has not yet been measured in bulge dwarfs, bulge giants in the \citet{mcwilliam:04} sample display a high [Eu/Fe] ratio.

\begin{figure}
\plotone{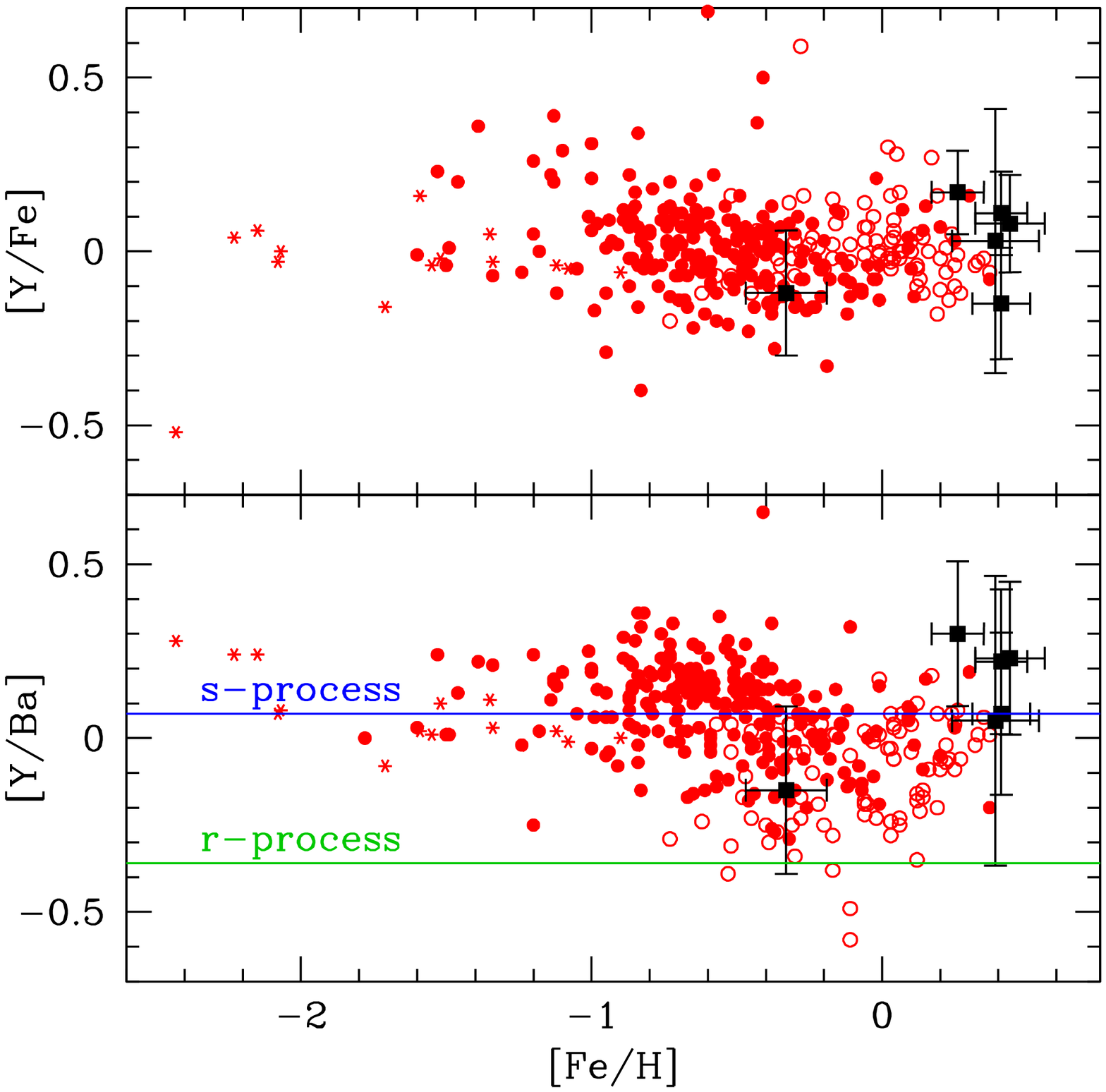}
\caption{\textit{Top:} Yttrium abundance for other six bulge bulge dwarfs from \citet{bensby209:09} and \citet{cohen:09} compared to stars in the thin disk (open circles), thick disk (closed circles), and halo (stars) from \citet{mashonkina:07}, \citet{bensby:05}, and \citet{reddy:06}. \textit{Bottom:} Ratio of light to heavy neutron capture elements exemplified by yttrium and barium, respectively. The lines mark the expected ratio from $s$-process only (blue) and $r$-process only (green) nucleosynthesis calculated by \citet{arlandini:99} for a star with [Fe/H$]=-0.30$ dex ($Z=\frac{1}{2}Z_\odot$)).}\label{fig:yttrium}
\end{figure}

The \textit{s}-process primarily occurs in AGB stars which should begin enriching the ISM on the same timescale as SNIa. Since the $\alpha$-element abundance indicates that OGLE-2007-BLG-514S formed after significant SNIa input, we would also expect to see an AGB component. The ratio of \textit{s}-process elements produced in AGB stars depends on the metallicity \citep{busso:01}. The \textit{s}-process has three major abundance peaks with Ba in the second peak. If we compare Ba with a first \textit{s}-process peak element like Y, we can determine the relative production.

We could not measure Y in OGLE-2007-BLG-514S, but Figure \ref{fig:yttrium} shows the Y abundance of the other six bulge dwarfs, which, except for the subgiant, show an overproduction of Y to Ba. The production of first \textit{s}-process peak elements peaks and exceeds that of Ba at high metallicity (see Fig 1 of \citealt{travaglio:04}). Perhaps metal-rich AGB stars could have formed and produced first \textit{s}-process peak elements (Y/Zr/Sr) over Ba resulting in the high [Y/Ba] ratio seen in \cite{cohen:09} and the low amount of Barium seen in OGLE-2007-BLG-514S. \citet{venn:04} explained the opposite phenomenon--a low [Y/Ba] ratio observed in Dwarf sphreoidal galaxies--with a similar argument; metal-poor AGB stars preferentially produce second \textit{s}-process peak elements.

The abundance of more neutron capture elements like Eu and Y in bulge stars would help clarify the origin of Ba abundances in the bulge.

\section{Conclusions}

We performed a detailed chemical abundance analysis of 15 elements in OGLE-2007-BLG-514S, a G dwarf star located in the bulge. Gravitational microlensing magnified this $I\sim21$ mag star by a factor $\gtrsim$1000, affording us a rare glimpse of the composition of such a faint star. This brings the number of dwarfs in the bulge with high-resolution spectra and published abundance ratios to eight.

OGLE-2007-BLG-514S boasts a high metallicity with [Fe/H]$=0.33\pm0.15$ dex, swelling the number of metal-rich bulge dwarfs to three-quarters of the sample. The prevalence of bulge dwarfs at high [Fe/H] conflicts with the wide distribution of bulge giant metallicities centered slightly below solar metallicity. We create a simple model for how a bulge metallicity gradient would affect the bulge giant MDF if the giants had the same spatial distribution as the dwarfs. A K-S test yields a 1.6\% probability that the bulge dwarf MDF is identical to our model giant MDF. We ascribe this incongruity to one of three possibilities. One option is that metal-rich dwarfs are not becoming metal-rich giants. \citet{cohen:08} proposes that this is a manifestation of the mechanism described in \citet{kilic:07} and \citet{kalirai:07}, namely that metal-rich dwarfs lose enough mass to skip the horizontal branch resulting in a comparatively small fraction of metal-rich giants. Otherwise, if bulge dwarfs and giants share the same underlying MDF, either a systematic offset between dwarfs and giants or small number statistics could produce an artificially low observed probability. Preliminary metallicities of six bulge dwarfs from the 2009 bulge season foreshadow a potential shift toward lower metallicity in the bulge dwarf MDF, which would bring it into closer agreement with the bulge giants \citep{bensbyIAU:09}.

Most elemental abundances confirm the trends observed in the other bulge dwarfs. The $\alpha$-elements follow the bulge giants' trend and display solar abundance ratios befitting their high metallicity. This indicates that SNIa enriched the ISM by the time OGLE-2007-BLG-514S formed, making it one of the younger stars in the bulge. The bulge dwarfs provide the first large sample of iron-peak abundances in the bulge. Although we lack high metallicity disk data, we note that the bulge [V/Fe] rises at high [Fe/H].

In [Na/Fe], we find a much tighter relationship among bulge dwarfs than bulge giants at [Fe/H]$\gtrsim 0$, despite claims by \citet{lecureur:07} that their bulge giants reflect the composition of the ISM at formation. While the bulge dwarf [Na/Fe] measurements could in principle be in error, we consider this unlikely given their small scatter and robustness relative to the solar abundance. One possibility is that giants experience enhanced extra mixing only in the metallicity region not yet sampled by the bulge dwarfs (eg. $0.0\lesssim\mathrm{[Fe/H]}\lesssim0.3$). Chemical evolution in the bulge could also produce both dwarfs and giants with high Na abundance in that metallicity region. Otherwise, enhanced extra mixing or measurement errors produced the giants' high [Na/Fe] and large scatter compared to the bulge dwarfs.

OGLE-2007-BLG-514S exhibits a strikingly low [Ba/Fe]$=-0.63\pm0.29$ dex in comparison to most other bulge dwarfs. Ba is typically produced in the $s$-process, although the $r$-process does generate small amounts. We propose that the $r$-process dominates in the bulge like it does in the halo. Alternatively, metal-rich AGB stars can overproduce first $s$-process peak elements relative to Ba. Although Ba represented the only measurable neutron-capture element in this star, future analyses of $r$-process and first $s$-process peak elements within the bulge can constrain which mechanism is responsible for the low Ba abundance in this star.

Gathering a larger sample of bulge dwarfs will improve the dwarf MDF and fill out abundance trends. The study of this important stellar population is made possible by microlensing. Bulge dwarfs, like OGLE-2007-BLG-514S, can serve as a valuable tool for probing the formation and chemical evolution of the bulge.

\acknowledgements
Our thanks to Michael Rauch who took spectra of this star using the Clay telescope. We would also like to thank Thomas Masseron for help with the abundance analysis. S.D. and A.G. were supported by NSF grant AST 0757888. The OGLE project is partially supported by the Polish MNiSW grant N20303032/4275 to AU.


\begin{thebibliography}{99}
\bibitem[Alonso et
al.(1996)]{alonso:96} Alonso, A., Arribas, S., \& Martinez-Roger, C.\ 1996, \aap, 313, 873

\bibitem[Alvarez
\& Plez(1998)]{alvarez:98} Alvarez, R., \& Plez, B.\ 1998, \aap, 330, 1109

\bibitem[Arlandini et al.(1999)]{arlandini:99} Arlandini, C.,
K{\"a}ppeler, F., Wisshak, K., Gallino, R., Lugaro, M., Busso, M.,
\& Straniero, O.\ 1999, \apj, 525, 886

\bibitem[Arnett(1971)]{arnett:71} Arnett, W.~D.\ 1971, \apj, 166,
153

\bibitem[Asplund et al.(2009)]{aspulund:09} Asplund, M., Grevesse,
N., Sauval, A.~J., \& Scott, P.\ 2009, arXiv:0909.0948

\bibitem[Baade(1946)]{baade:46} Baade, W.\ 1946, \pasp, 58, 249

\bibitem[Baade(1958)]{baade:58} Baade, W.\ 1958, Ricerche
Astronomiche, 5, 303

\bibitem[Bedin et al.(2005)]{bedin:05} Bedin, L.~R., Salaris,
M., Piotto, G., King, I.~R., Anderson, J., Cassisi, S.,
\& Momany, Y.\ 2005, \apjl, 624, L45

\bibitem[Bedin et al.(2008)]{bedin:08} Bedin, L.~R., Salaris,
M., Piotto, G., Cassisi, S., Milone, A.~P., Anderson, J.,
\& King, I.~R.\ 2008, \apjl, 679, L29

\bibitem[Bennett et al.(2008)]{bennett:08} Bennett, D.~P., et al.\
2008, \apj, 684, 663

\bibitem[Bensby et
al.(2003)]{bensby:03} Bensby, T., Feltzing, S., \& Lundstr{\"o}m, I.\ 2003, \aap, 410, 527

\bibitem[Bensby et
al.(2004)]{bensby:04} Bensby, T., Feltzing, S., \& Lundstr{\"o}m, I.\ 2004, \aap, 415, 155

\bibitem[Bensby et
al.(2005)]{bensby:05} Bensby, T., Feltzing, S., Lundstr{\"o}m, I., \& Ilyin, I.\ 2005, \aap, 433, 185

\bibitem[Bensby et al.(2009a)]{bensby209:09} Bensby, T., et al.\
2009, arXiv:0903.3044

\bibitem[Bensby et al.(2009b)]{bensby76:09} Bensby, T., et al.\
2009, arXiv:0906.2235

\bibitem[Bensby et al.(2009c)]{bensbyIAU:09} Bensby, T., et al.\
2009, arXiv:0908.2779

\bibitem[Bernstein et al.(2003)]{bernstein:03} Bernstein, R., Shectman, S.~A., Gunnels, S.~M., Mochnacki, S., \& Athey, A.~E.\ 2003, \procspie, 4841, 1694

\bibitem[Busso et al.(2001)]{busso:01} Busso, M., Marengo, M.,
Travaglio, C., Corcione, L.,
\& Silvestro, G.\ 2001, Memorie della Societa Astronomica Italiana, 72, 309

\bibitem[Cochrane et al.(1998)]{cochrane:98} Cochrane, E.~C.~A.,
Benton, D.~M., Forest, D.~H.,
\& Griffith, J.~A.~R.\ 1998, Journal of Physics B Atomic Molecular Physics, 31, 2203

\bibitem[Castellani \& Castellani(1993)]{castellani:93} Castellani, M., \& Castellani, V.\ 1993, \apj, 407, 649

\bibitem[Castelli
\& Kurucz(2003)]{castelli:03} Castelli, F., \& Kurucz, R.~L.\ 2003, Modelling of Stellar Atmospheres, 210, 20P

\bibitem[Cavallo et al.(2003)]{cavallo:03} Cavallo, R.~M., Cook,
K.~H., Minniti, D., \& Vandehei, T.\ 2003, \procspie, 4834, 66

\bibitem[Chen et al.(2003)]{chen:03} Chen, Y.~Q., Zhao, G.,
Nissen, P.~E., Bai, G.~S., \& Qiu, H.~M.\ 2003, \apj, 591, 925

\bibitem[Childs et al.(1979)]{childs:79} Childs, W.~J., Poulsen,
O., Goodman, L.~S., \& Crosswhite, H.\ 1979, \pra, 19, 168

\bibitem[Cohen et al.(2008)]{cohen:08} Cohen, J.~G., Huang, W.,
Udalski, A., Gould, A., \& Johnson, J.~A.\ 2008, \apj, 682, 1029

\bibitem[Cohen et al.(2009)]{cohen:09} Cohen, J.~G., Thompson,
I.~B., Sumi, T., Bond, I., Gould, A., Johnson, J.~A., Huang, W.,
\& Burley, G.\ 2009, arXiv:0904.2020

\bibitem[Cunha
\& Smith(2006)]{cunha:06} Cunha, K., \& Smith, V.~V.\ 2006, \apj, 651, 491

\bibitem[D'Cruz et al.(1996)]{dcruz:96} D'Cruz, N.~L., Dorman,
B., Rood, R.~T., \& O'Connell, R.~W.\ 1996, \apj, 466, 359

\bibitem[Dehnen
\& Binney(1998)]{dehnen:98} Dehnen, W., \& Binney, J.~J.\ 1998, \mnras, 298, 387

\bibitem[Demarque et al.(2004)]{demarque:04} Demarque, P., Woo,
J.-H., Kim, Y.-C., \& Yi, S.~K.\ 2004, \apjs, 155, 667

\bibitem[Denissenkov et al.(2006)]{denissenkov:06} Denissenkov, P.~A.,
Pinsonneault, M., \& Terndrup, D.~M.\ 2006, \apj, 651, 438

\bibitem[Feltzing et
al.(2007)]{feltzing:07} Feltzing, S., Fohlman, M., \& Bensby, T.\ 2007, \aap, 467, 665

\bibitem[Feltzing
\& Gilmore(2000)]{feltzing:00} Feltzing, S., \& Gilmore, G.\ 2000, \aap, 355, 949

\bibitem[Fulbright et al.(2006)]{fulbright:06} Fulbright, J.~P.,
McWilliam, A., \& Rich, R.~M.\ 2006, \apj, 636, 821

\bibitem[Fulbright et al.(2007)]{fulbright:07} Fulbright, J.~P.,
McWilliam, A., \& Rich, R.~M.\ 2007, \apj, 661, 1152

\bibitem[Gould(2000)]{gould:00} Gould, A.\ 2000, \apj, 535, 928

\bibitem[Gratton(1989)]{gratton:89} Gratton, R.~G.\ 1989, \aap, 208, 171

\bibitem[Grevesse
\& Sauval(1998)]{grevsauv} Grevesse, N., \& Sauval, A.~J.\ 1998, Space Science Reviews, 85, 161

\bibitem[Hansen(2005)]{hansen:05} Hansen, B.~M.~S.\ 2005, \apj,
635, 522

\bibitem[Holmberg et al.(2006)]{holmberg:06} Holmberg, J., Flynn,
C., \& Portinari, L.\ 2006, \mnras, 367, 449

\bibitem[Howard et al.(2008)]{howard:08} Howard, C.~D., Rich,
R.~M., Reitzel, D.~B., Koch, A., De Propris, R.,
\& Zhao, H.\ 2008, \apj, 688, 1060

\bibitem[Ivans et al.(1999)]{ivans:99} Ivans, I.~I., Sneden, C.,
Kraft, R.~P., Suntzeff, N.~B., Smith, V.~V., Langer, G.~E.,
\& Fulbright, J.~P.\ 1999, \aj, 118, 1273

\bibitem[Janes
\& Kassas(1997)]{janes:97} Janes, K., \& Kassas, M.\ 1997, The Third Pacific Rim Conference on Recent Development on Binary Star Research, 130, 107

\bibitem[Johnson et al.(2009)]{johnson:09} Johnson, J.~A.,
Dong, Subo, \& Gould, A.\ 2009

\bibitem[Johnson et al.(2007)]{johnson:07} Johnson, J.~A.,
Gal-Yam, A., Leonard, D.~C., Simon, J.~D., Udalski, A.,
\& Gould, A.\ 2007, \apjl, 655, L33

\bibitem[Johnson et al.(2008)]{johnson:08} Johnson, J.~A., Gaudi,
B.~S., Sumi, T., Bond, I.~A., \& Gould, A.\ 2008, \apj, 685, 508

\bibitem[Johnson et al.(2006)]{johnson:06} Johnson, J.~A., Ivans,
I.~I., \& Stetson, P.~B.\ 2006, \apj, 640, 801

\bibitem[Kalirai et al.(2007)]{kalirai:07} Kalirai, J.~S.,
Bergeron, P., Hansen, B.~M.~S., Kelson, D.~D., Reitzel, D.~B., Rich, R.~M.,
\& Richer, H.~B.\ 2007, \apj, 671, 748

\bibitem[Kane
\& Sahu(2006)]{kane:06} Kane, S.~R., \& Sahu, K.~C.\ 2006, \apj, 637, 752

\bibitem[Kilic et al.(2007)]{kilic:07} Kilic, M., Stanek, K.~Z.,
\& Pinsonneault, M.~H.\ 2007, \apj, 671, 761

\bibitem[Kormendy
\& Kennicutt(2004)]{kormendy:04} Kormendy, J., \& Kennicutt, R.~C., Jr.\ 2004, \araa, 42, 603

\bibitem[Kraft(1994)]{kraft:94} Kraft, R.~P.\ 1994, \pasp, 106,
553

\bibitem[Lef\`{e}bvre et al.(2002)]{lefebvre:02} Lef\`{e}bvre, P.-H.,
Garnir, H.-P., \& Bi{\'e}mont, E.\ 2002, \physscr, 66, 363

\bibitem[Lecureur et
al.(2007)]{lecureur:07} Lecureur, A., Hill, V., Zoccali, M., Barbuy, B., G{\'o}mez, A., Minniti, D., Ortolani, S., \& Renzini, A.\ 2007, \aap, 465, 799

\bibitem[Lennon et al.(1996)]{lennon:96} Lennon, D.~J., Mao, S.,
Fuhrmann, K., \& Gehren, T.\ 1996, \apjl, 471, L23

\bibitem[Luck
\& Heiter(2007)]{luck:07} Luck, R.~E., \& Heiter, U.\ 2007, \aj, 133, 2464

\bibitem[Mashonkina et al.(2007)]{mashonkina:07} Mashonkina, L.~I.,
Vinogradova, A.~B., Ptitsyn, D.~A., Khokhlova, V.~S.,
\& Chernetsova, T.~A.\ 2007, Astronomy Reports, 51, 903

\bibitem[Mashonkina
\& Zhao(2006)]{mashonkina:06} Mashonkina, L., \& Zhao, G.\ 2006, \aap, 456, 313

\bibitem[Matteucci
\& Brocato(1990)]{matteucci:90} Matteucci, F., \& Brocato, E.\ 1990, \apj, 365, 539

\bibitem[Matteucci et al.(2009)]{matteucci:09} Matteucci, F.,
Spitoni, E., Recchi, S., \& Valiante, R.\ 2009, arXiv:0905.0272

\bibitem[McWilliam et al.(2008)]{mcwilliam:08} McWilliam, A.,
Matteucci, F., Ballero, S., Rich, R.~M., Fulbright, J.~P.,
\& Cescutti, G.\ 2008, \aj, 136, 367

\bibitem[McWilliam et al.(1995)]{mcwilliam:95} McWilliam, A.,
Preston, G.~W., Sneden, C., \& Searle, L.\ 1995, \aj, 109, 2757

\bibitem[McWilliam
\& Rich(1994)]{mcwilliam:94} McWilliam, A., \& Rich, R.~M.\ 1994, \apjs, 91, 749

\bibitem[McWilliam
\& Rich(2004)]{mcwilliam:04} McWilliam, A., \& Rich, R.~M.\ 2004, Origin and Evolution of the Elements

\bibitem[Mel{\'e}ndez et
al.(2008)]{melendez:08} Mel{\'e}ndez, J., et al.\ 2008, \aap, 484, L21

\bibitem[Minniti et al.(1998)]{minniti:98} Minniti, D., Vandehei,
T., Cook, K.~H., Griest, K., \& Alcock, C.\ 1998, \apjl, 499, L175

\bibitem[Nair
\& Miralda-Escud{\'e}(1999)]{nair:99} Nair, V., \& Miralda-Escud{\'e}, J.\ 1999, \apj, 515, 206

\bibitem[Ortolani et al.(1995)]{ortolani:95} Ortolani, S., Renzini,
A., Gilmozzi, R., Marconi, G., Barbuy, B., Bica, E.,
\& Rich, R.~M.\ 1995, \nat, 377, 701

\bibitem[Palmeri et al.(1995)]{palmeri:95} Palmeri, P., Biemont,
E., Aboussaid, A.,
\& Godefroid, M.\ 1995, Journal of Physics B Atomic Molecular Physics, 28, 3741

\bibitem[Ram{\'{\i}}rez
\& Mel{\'e}ndez(2005)]{ramirez:05} Ram{\'{\i}}rez, I., \& Mel{\'e}ndez, J.\ 2005, \apj, 626, 465

\bibitem[Reddy et al.(2006)]{reddy:06} Reddy, B.~E., Lambert,
D.~L., \& Allende Prieto, C.\ 2006, \mnras, 367, 1329

\bibitem[Reddy et al.(2003)]{reddy:03} Reddy, B.~E., Tomkin, J.,
Lambert, D.~L., \& Allende Prieto, C.\ 2003, \mnras, 340, 304

\bibitem[Rich \& Origlia (2005)]{rich:05} Rich, R.~M., \& Origlia, L.\ 2005, \apj, 634, 1293

\bibitem[Rich et al.(2007)]{rich:07} Rich, R.~M., Origlia, L., \& Valenti, E.\ 2007, \apjl, 665, L119

\bibitem[Ryde et al.(2009)]{ryde:09} Ryde, N., Edvardsson, B.,
Gustafsson, B., Eriksson, K., Kaufl, H.~U., Siebenmorgen, R.,
\& Smette, A.\ 2009, arXiv:0902.2124

\bibitem[Smith et al.(2005)]{smith:05} Smith, V.~V., Cunha, K.,
Ivans, I.~I., Lattanzio, J.~C., Campbell, S.,
\& Hinkle, K.~H.\ 2005, \apj, 633, 392

\bibitem[Sneden et al.(2004)]{sneden:04} Sneden, C., Kraft,
R.~P., Guhathakurta, P., Peterson, R.~C.,
\& Fulbright, J.~P.\ 2004, \aj, 127, 2162

\bibitem[Stanek et al.(2000)]{stanek:00} Stanek, K.~Z., Kaluzny,
J., Wysocka, A., \& Thompson, I.\ 2000, Acta Astronomica, 50, 191

\bibitem[Sumi(2004)]{sumi:04} Sumi, T.\ 2004, \mnras, 349, 193

\bibitem[Tinsley(1979)]{tinsley:79} Tinsley, B.~M.\ 1979, \apj,
229, 1046

\bibitem[Travaglio et al.(2004)]{travaglio:04} Travaglio, C.,
Gallino, R., Arnone, E., Cowan, J., Jordan, F.,
\& Sneden, C.\ 2004, \apj, 601, 864

\bibitem[van Loon et al.(2008)]{vanloon:08} van Loon, J.~T.,
Boyer, M.~L., \& McDonald, I.\ 2008, \apjl, 680, L49

\bibitem[Venn et al.(2004)]{venn:04} Venn, K.~A., Irwin, M.,
Shetrone, M.~D., Tout, C.~A., Hill, V.,
\& Tolstoy, E.\ 2004, \aj, 128, 1177

\bibitem[Woosley
\& Weaver(1995)]{woosley:95} Woosley, S.~E., \& Weaver, T.~A.\ 1995, \apjs, 101, 181

\bibitem[Yi et al.(2001)]{yi:01} Yi, S., Demarque, P., Kim,
Y.-C., Lee, Y.-W., Ree, C.~H., Lejeune, T.,
\& Barnes, S.\ 2001, \apjs, 136, 417

\bibitem[Yoo et al.(2004)]{yoo:04} Yoo, J., et al.\ 2004,
\apj, 603, 139

\bibitem[Zhang et
al.(2006)]{zhang:06} Zhang, H.~W., Gehren, T., Butler, K., Shi, J.~R., \& Zhao, G.\ 2006, \aap, 457, 645

\bibitem[Zoccali et al.(2003)]{zoccali:03} Zoccali, M., et al.\ 2003, \aap, 399, 931

\bibitem[Zoccali et al.(2008)]{zoccali:08} Zoccali, M., Hill, V., Lecureur, A., Barbuy, B., Renzini, A., Minniti, D., G{\'o}mez, A., \& Ortolani, S.\ 2008, \aap, 486, 177

\end{thebibliography}
\end{document}